# Vertically-oriented MoS$_2$ nanosheets for nonlinear optical devices

M. Bolhuis[1], J. Hernandez-Rueda[1], S. E. van Heijst[1], M. Tinoco Rivas[1,†], L. Kuipers[1], and S. Conesa-Boj[1,*]

[1] Kavli Institute of Nanoscience, Delft University of Technology, 2628CJ Delft, The Netherlands

**ABSTRACT**

Transition metal dichalcogenides such as $MoS_2$ represent promising candidates for building blocks of ultra-thin nanophotonic devices. For such applications, vertically-oriented $MoS_2$ (v-$MoS_2$) nanosheets could be advantageous as compared to conventional horizontal $MoS_2$ (h-$MoS_2$) given that their inherent broken symmetry would favor an enhanced nonlinear response. However, the current lack of a controllable and reproducible fabrication strategy for v-$MoS_2$ limits the exploration of this potential. Here we present a systematic study of the growth of v-$MoS_2$ nanosheets based on the sulfurization of a pre-deposited Mo-metal seed layer. We demonstrate that the sulfurization process at high temperatures is driven by the diffusion of Sulfur from the vapor-solid interface to the Mo seed layer. Furthermore, we verify an enhanced nonlinear response in the resulting v-$MoS_2$ nanostructures as compared to their horizontal counterparts. Our results represent a stepping stone towards the fabrication of low-dimensional TMD-based nanostructures for versatile nonlinear nanophotonic devices.

**INTRODUCTION**

Two-dimensional (2D) materials such as transition metal dichalcogenides (TMDs) have been extensively exploited for a wide range of applications including optoelectronics devices[1,2] and catalysis[3] among others. Specifically, these materials exhibit numerous remarkable electronic and optical properties thanks to their broken inversion symmetry.[4,5,6] Significant attention has been recently devoted to their nonlinear optical response,[7] which makes TMDs ideal building blocks for ultra-thin[8] nonlinear photonic devices.[9]

Such nonlinear optical effects have been demonstrated in horizontal $MoS_2$ (h-$MoS_2$) monolayers, displaying a marked dependence on the specific crystalline symmetry and orientation.[10] For instance, a nonlinear optical response has been reported at the atomic edges of h-$MoS_2$ crystals, where translation symmetry is broken.[11] These findings suggest that vertically-oriented $MoS_2$ (v-$MoS_2$) nanosheets, a configuration that maximizes the number of exposed edge sites, could represent a promising platform to enhance the second-order nonlinear response and realize a novel candidate for the building blocks of high efficiency nanophotonic devices.

On the one hand, significant progress has been achieved in the understanding of the growth dynamics of horizontal $MoS_2$. Different fabrication methods have been employed, such as chemical vapor deposition (CVD) techniques,[12,13] the direct sulfurization of a pre-deposited molybdenum (Mo) seed layer,[14,15] and by using a vapor phase reaction with $MoO_3$.[16,17]

On the other hand, the growth mechanism of its vertical counterpart, v-$MoS_2$, remains still poorly understood. Several attempts at explaining the growth of vertically-oriented $MoS_2$ nanostructures have been put forward. For instance, in the context of growth strategies based on the sulfurization of a pre-existing Mo-metal layer,[18] it has been shown that a low reaction temperature of 550 °C results into v-MoS$_2$ with the kinetically-controlled growth being diffusion limited. Furthermore, it has been reported that the orientation of the resulting MoS$_2$ layers with respect to the substrate is sensitive to the thickness of the Mo-metal layer.[19] Specifically, thicker and more uniform Mo-metal seed layers lead to a higher fraction of v-MoS$_2$ layers. Additionally, theoretical models have been also constructed aiming to describe the synthesis of vertically-oriented MoS$_2$ based on the solid-vapor reaction,[18,19,20,21] though most of these predictions remain to be verified.

Given this state of affairs, achieving further progress towards a controllable and reproducible fabrication strategy for v-MoS$_2$ requires detailed studies of the associated growth mechanisms. Here we present a systematic investigation of the growth mechanism of vertical MoS$_2$ nanosheets based on the sulfurization of a pre-deposited Mo-metal seed layer. Thanks to an extensive structural cross-section characterization by means of transmission electron microscopy (TEM), we demonstrate that the sulfurization mechanism for temperatures between 600 and 700 °C proceeds via diffusion. These findings imply that during the Sulfur reaction the growth propagates from the vapor-solid interface inwards into the Mo seed layer. In addition, we investigate the prospects of the resulting v-MoS$_2$ nanostructures for nonlinear optical applications. We verify an enhanced nonlinear response as compared to their h-MoS$_2$ counterparts, confirmed by the observation of second-harmonic generation and sum-frequency generation.

Our results provide a stepping stone towards the large-scale fabrication of high-quality v-MoS$_2$ nanosheets, and represent a crucial step in a program aimed at designing and fabricating low-dimensional nanostructures for the development of efficient and versatile nonlinear optical devices based on 2D materials.

**RESULTS and DISCUSSION**

A two-step process was used for synthesizing the vertically-oriented $MoS_2$ nanosheets by means of the sulfurization of a pre-deposited Mo-metal layer. A thick Mo seed layer of 700 nm was chosen in order to be able to ascertain the dependence on the reaction time and the sulfurization depth. The sulfurization process was carried out inside a three-zone hot-wall horizontal tube. Before and during the sulfurization, an Argon gas flow was used to prevent any possible oxidation as well as a carrier to transport the Sulphur vapor phase to the substrate. Further details about the growth process are described in the Supporting Information (SI).

The orientation of the resulting $MoS_2$ nanosheets was investigated as a function of the reaction temperature. For these studies, we considered several growth temperatures in the range between 500 °C and 700 °C. We now highlight the results obtained for the two limiting cases corresponding to temperatures of 500 °C and 700 °C.

**Figures 1a** and **1b** display top-view scanning electron microscopy (SEM) images of the sulfurized Mo-metal layer at 500 and 700 °C respectively. One observes marked differences in the morphology between the results obtained in the two growths. While the vertical nature of the $MoS_2$ nanosheets grown at a temperature of 500 °C is unambiguous from the SEM image (**Figure 1a**), the same inspection is less conclusive for the sample grown at 700 °C, whose surface exhibits a granular-like aspect (**Figure 1b**).

**Figures 1c** and **1d** depict Raman spectroscopy measurements taken in the samples grown at 500 and 700°C, respectively. These Raman spectra are dominated by the in-plane $E^1_{2g}$ and the out-of-plane $A_{1g}$ Raman modes. The appearance and position of these peaks is consistent with a trigonal prismatic (2H-MoS$_2$) crystal phase, further confirming the successful MoS$_2$ growth.[22]

Interestingly, we find that for the sample grown at 500 (700) °C the ratio of intensities between the $A_{1g}$ and $E^1_{2g}$ peaks increases by a factor 2 (3) as compared to regular MoS$_2$ flakes, where the two peaks exhibit comparable intensities.[19] Given that the $A_{1g}$ and $E^1_{2g}$ Raman peaks are associated respectively with the out-of-plane and in-plane vibration modes, these results suggest that our specimens exhibit the presence of v-MoS$_2$ nanosheets in the samples grown at both temperatures, therefore indicating a higher density of exposed edges.

Furthermore, in **Figure 1c (1d)** a third peak located at 284 (283) cm$^{-1}$ and associated to the $E_{1g}$ Raman mode is also observed. This mode is forbidden in backscattering experiments,[23] which implies that it should not be observed when the incident laser beam is perpendicular to the basal plane, as happens for horizontal MoS$_2$. The presence of the $E_{1g}$ mode thus indicates that the laser beam is no longer perpendicular to the incident (basal) plane, providing a further confirmation of the presence of vertically-oriented (with respect to the substrate) MoS$_2$ nanosheets in both samples.

To investigate the orientation of the grown MoS$_2$, we further complement this surface analysis with a cross-section study performed by focus ion beam (FIB) followed by SEM inspection. For the sample grown at 500 °C (**Figure 2a**), the thickness of the original Mo seed layer (700 nm) remains mostly unaffected after sulfurization. However, for the sample grown at 700 °C (**Figure 2b**), the Mo seed layer is close to being fully sulfurized. From the cross-section SEM image, one observes that the original thickness of the Mo-metal layer has been reduced down to a length of 200 nm, with the rest of the Mo layer sulfurized into $MoS_2$. In this specific case, the final sulfurized layer has a length of 1.95 µm, representing a factor of around three increase as compared to the thickness of the initial Mo layer.

From these experimental results, one can therefore distinguish two distinctive dynamics for the growth of the v-$MoS_2$ nanosheets. On the one hand, at 500 °C, the Sulfur only reacts on the immediate Mo seed layer surface, leading to vertically-standing $MoS_2$ as can be observed in the top-view SEM image in **Figure 1a**. On the other hand, at 700 °C, the Sulfur diffuses through the Mo seed layer by consuming it and forming v-$MoS_2$ layers. To investigate the crystalline quality of the vertical $MoS_2$ grown at 700 °C, we have produced a cross-section sample using FIB and then analyzed it by means of transmission electron microscopy (TEM). The three different contrasts observed in **Figure 3a** reveal the sequence $MoS_2$, Mo-metal layer, and Silicon. The high-resolution TEM (HRTEM) measurements performed at the $MoS_2$/Mo interface (**Figure 3b**) indicate that the $MoS_2$ grows vertically with respect to the Mo seed layer. The distance between two neighboring $MoS_2$ layers was measured to be about 0.65 nm (**Figure 3c**), consistent with previous results in the literature.[24] Energy dispersive X-ray spectroscopy (EDX) measurements along the length of the whole cross-section (**Figure 3d**) provide clear evidence of the Sulfur diffusion into the Mo seed layer that results into the growth of the vertically-oriented $MoS_2$ nanosheets.

*Growth mechanism of vertically-oriented $MoS_2$ nanosheets at 700 °C*

In order to further elucidate the sulfurization mechanism of the v-$MoS_2$ at 700 °C, we have sulfurized for different times sample containing each 700 nm of Mo seed layer.

Subsequently, we prepared with FIB cross-sections lamellas for TEM and EDX inspection that allowed us to determine the thicknesses of the MoS$_2$ layer and of the consumed Mo. We have only considered reactions times equal to and larger than 15 min, in a way that the Mo seed layer will be always completely sulfurized, see **Figure S5**. In **Figure 4a** we display the value of the consumed Mo seed layer as a function of the reaction time. These measurements are fitted to a model of the form $z = K(t - t_0)^n$, with $z$ and $t$ are the thickness of the consumed Mo seed layer and the reaction time respectively. The best-fit value for the growth exponent *n* is found to be 0.48. The fact that *n* is very close to 1/2 is consistent with a sulfurization process dominated by the diffusion mechanism.

Since the growth exponent is essentially 0.5, we can use the relation $z_{dif} = 2\sqrt{Dt}$ to extract from the data the diffusion coefficient of Sulfur within the Mo seed layer. The best-fit value for *D* is calculated to be 20.7 nm$^2$/s, similar to the diffusion coefficients measured for Sulfur in other metals.[25] Our findings therefore confirm that the reaction is driven by the diffusion of the Sulfur and that the consumption of Sulfur takes place predominantly at the boundary between the Mo seed layer and the grown v-MoS$_2$ layers. These results are consistent with theoretical models of the diffusion-reaction growth proposed for the synthesis of vertically-oriented MoS$_2$.[26]

From the HR-TEM cross-section lamella analysis (**Figure S4**) we also found an orientation-disordered region extending from the surface to the first 20 nanometers, where both vertical and horizontal MoS$_2$ nanosheets are present. The reaction mechanism in this initial region appears to be self-limited,[26] but then from this point onwards the growth front results into v-MoS$_2$ being dominated by the diffusion kinetics of Sulfur, as discussed above.

The influence of the temperature on the rate of the consumed Mo seed layer was also examined in the range between 600 and 700 °C. The lower range of this interval corresponds to the minimum temperature required to initiate the diffusion of the Sulfur within the Mo seed layer.[18] A fixed reaction time of 30 min was adopted in these experiments. **Figure 4b** highlights the effect of the rate of the $MoS_2$ thickness growth as a function of the reaction temperature. The experimental Arrhenius plot can be fitted very well by a straight line to determine the activation energy $E_A$, which turns out to be 192.44 kJ/mol. These findings provide additional evidence that the Sulfur diffuses in the range between 600 and 700 °C through the Mo seed layer, leading to the phase transformation into vertically-oriented $MoS_2$ nanosheets.

*Nonlinear optical effects in vertical-oriented $MoS_2$ nanosheets.*

As mentioned above, TMD materials have generated ample attention because of their nonlinear optical response.[7] In particular, second and third-order nonlinear optical processes have recently been demonstrated in few-layer $MoS_2$ flakes. Second-harmonic (SHG) and sum-frequency generation (SFG) have been shown to be more efficiently generated in thin $MoS_2$ flakes with a few atomic layers. [27,28] The intensity of both processes exponentially increases when decreasing the number of layers, thus revealing $MoS_2$ monolayers to be the most efficient thickness in order to generate second-order processes. Moreover, due to inversion symmetry breaking, only odd-layered $MoS_2$ flakes present second-order processes. In contrast, third-order processes, such as four-wave mixing (FWM), gradually increase their intensity with increasing number of layers (reaching saturation for a certain thickness) irrespective of their parity.

Here we use multiphoton spectroscopy to explore the nonlinear optical response of the synthesized vertical MoS$_2$ nanosheets reported in this work, and compare the results with those of their h-MoS$_2$ counterparts. **Figure 5a** depicts the spectra of the nonlinear emission when excited with a laser pulse at 776 nm (blue line) and with two synchronized laser pulses at 776 nm and 1210 nm (red line). The labels indicate the nonlinear mechanisms that originate the emission at each spectral peak. The spectra in **Figure 5a** illustrate all the above-mentioned processes. In order to benchmark the response of the v-MoS$_2$ nanosheets in **Figure 5a**, we also measured the nonlinear emission of the horizontally-oriented case. **Figure 5b** shows the spectra collected upon laser illumination with a single 776 nm beam (blue line) and both beams at 776 nm and 1210 nm (red line). Although both MoS$_2$ geometries exhibit the same emission peaks (except SHG$_2$), it is clear that vertically-oriented MoS$_2$ nanosheets favor second-order processes (*i.e.* SFG in **Figure 5a**) over third-order processes (*i.e.* THG and FWM in **Figure 5a**). In contrast, the horizontal configuration shows the opposite trend where third-order processes dominate and second-order processes are three orders of magnitude smaller due to inversion symmetry, in agreement with previous results in the literature.[10,29]

By using the intensity of the spectral peaks for SFG and SHG along with the power of the excitation beams, one can estimate the relative ratio of the nonlinear susceptibility.[7,30] The second-order susceptibility ratios between the vertical and horizontal geometries extracted from SHG and SFG turn out to be $\chi_V^{(2)}/\chi_H^{(2)}|_{SHG} = 22.9 \pm 1.5$ and $\chi_V^{(2)}/\chi_H^{(2)}|_{SFG} = 24.7 \pm 1.0$, respectively. These results indicate that vertically-oriented MoS$_2$ nanosheets significantly favor second-order nonlinear processes as compared to their horizontal counterparts. As mentioned above, this enhancement can be traced back to the effects of inversion symmetry breaking at the exposed edges of the nanosheets.

**CONCLUSIONS**

In this work we have reported the controllable and reproducible fabrication of vertically-oriented $MoS_2$ nanosheets. We have demonstrated that the phase transformation from Mo seed layers to vertically-oriented $2H-MoS_2$ nanosheets can be achieved by means of reacting the pre-deposited Mo-metal layer with Sulfur at relatively high temperatures, in the range between 600 and 700 °C. Following a systematic characterization analysis of these v-$MoS_2$ nanosheets using TEM and EDX, we have established that in this range of temperatures the sulfurization mechanism proceeds via diffusion.

Furthermore, we have investigated the nonlinear optical response of the resulting v-$MoS_2$ nanostructures, including second-harmonic generation, and found an enhanced second-order nonlinear response as compared to the h-$MoS_2$ case. The latter property could be explained by the effects of inversion symmetry breaking at the exposed edges of the nanosheets. Our findings therefore represent a stepping stone towards the fabrication of low-dimensional TMD-based nanostructures for versatile nonlinear nanophotonic devices.

## METHODS

**Sample preparation.** A two-step process was used for synthesizing the vertically-oriented $MoS_2$ nanosheets. First, a 700 nm-thick Mo seed layer was pre-deposited on a $Si/SiO_2$ wafer using magnetron sputtering. The sulfurization was carried out in a gradient tube furnace from Carbolite Gero. Note that Argon gas was used as a carrier gas. The Ar flow was set up 150 sccm for all the syntheses. This Mo seed layer was placed in the middle zone and gradually heated up to the reaction temperature. Once the sample reached the reaction temperature, 400 mg of Sulfur was heated to 220 °C. The Sulfur was placed upstream from the sample. Further details on the synthesis can be found in the SI.

**Characterization techniques.**

*Transmission electron microscopy (TEM) measurements.* TEM and energy-dispersive X-ray spectroscopy were carried out in a Titan Cube microscope operated at 300 kV. Its spatial resolution at Scherzer defocus conditions is 0.08 nm in the High-resolution TEM mode, whilst the resolution is around 0.19 nm in the High-Angle Annular Dark-Field Scanning Transmission Electron Microscopy (HAADF-STEM). Cross-section TEM lamellas were fabricated using a FEI Helios G4 CX™ dual beam system.

*Raman Spectroscopy.* Raman spectroscopy was performed using a Renishaw InVia Reflex™ confocal Raman microscope. The wavelength of the exiting laser was 514 nm, and a 1800 l/mm grating was used resulting in a spectral resolution of around 1 cm$^{-1}$.

*Optical measurements.* The specimen was illuminated with two synchronized laser pulses at $\lambda_1$ = 776 nm and $\lambda_2$ = 1210 nm that were temporally and spatially overlapped. The laser pulses were generated by a femtosecond laser oscillator (Tsunami, Spectra-Physics) and an optical parametric oscillator (OPAL, Spectra-Physics). Both laser beams were focused onto the sample using a microscope objective (Olympus UP- LSAPO 40x/0.95), which also collects the emitted light originated through the nonlinear laser-v-MoS2-nanosheets interaction. The collected light was filtered and imaged onto the slit of a spectrometer (PI, Spectra Pro 2300I). In these measurements the second-order susceptibility $\chi^{(2)}$ gives rise to the second-harmonic generation (SHG) and the sum-frequency generation (SFG) at $2\omega_1$, $2\omega_2$ and $\omega_1+\omega_2$. The third-order susceptibility $\chi^{(3)}$ mediates the generation of nonlinear polarization $P^{(3)}$, resulting in third-harmonic generation (THG) and four-wave mixing (FWM) at $3\omega_2$ and $2\omega_1-\omega_2$, respectively.

**FIGURES**

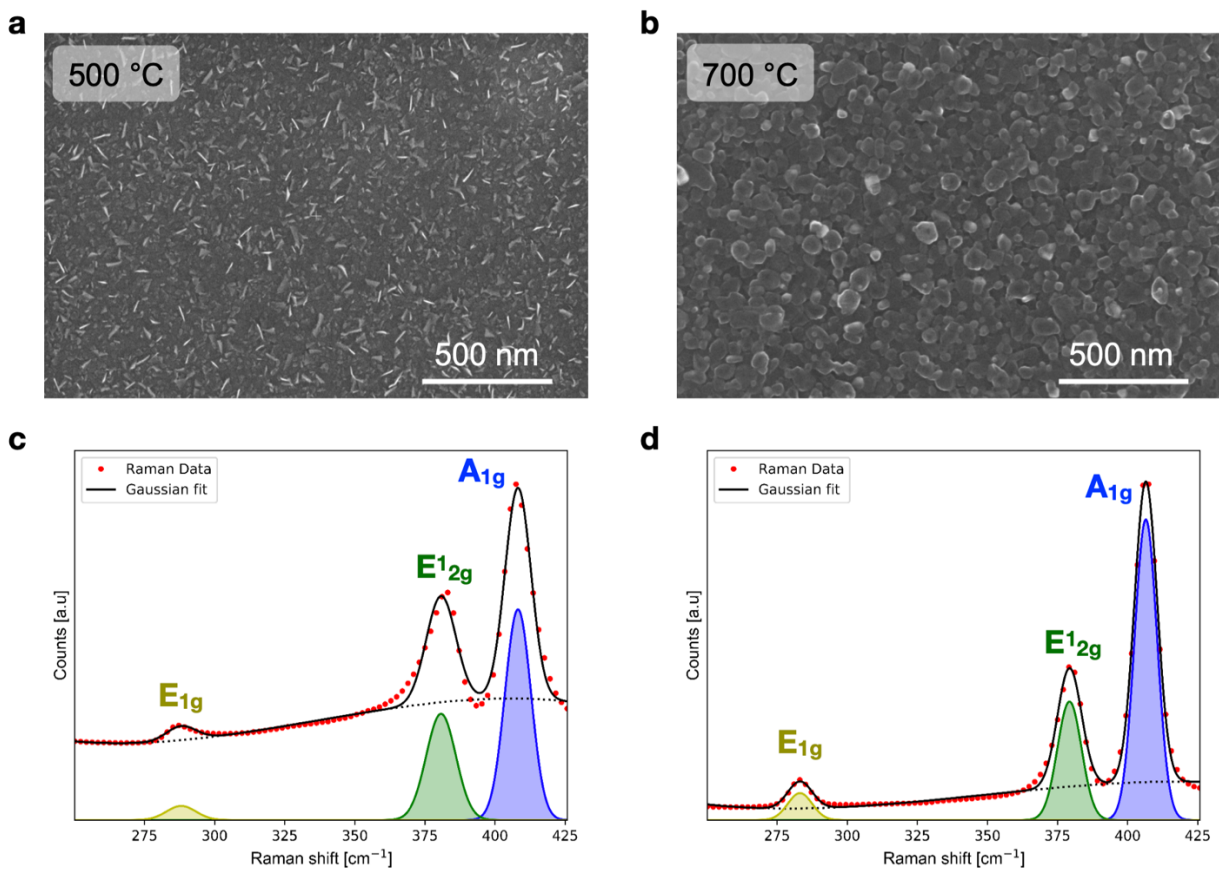

**Figure 1. (a)** and **(b)** Top-view SEM images of the surface of the samples sulfurized at 500 and 700 °C, respectively. **(c)** and **(d)** The associated Raman spectra, where the two main Raman modes ($A_{1g}$ and $E^1_{2g}$) are visible along with the smaller $E_{1g}$ mode. The similarities between the two spectra suggest the vertical orientation of the $MoS_2$ nanosheets in the samples grown at both temperatures.

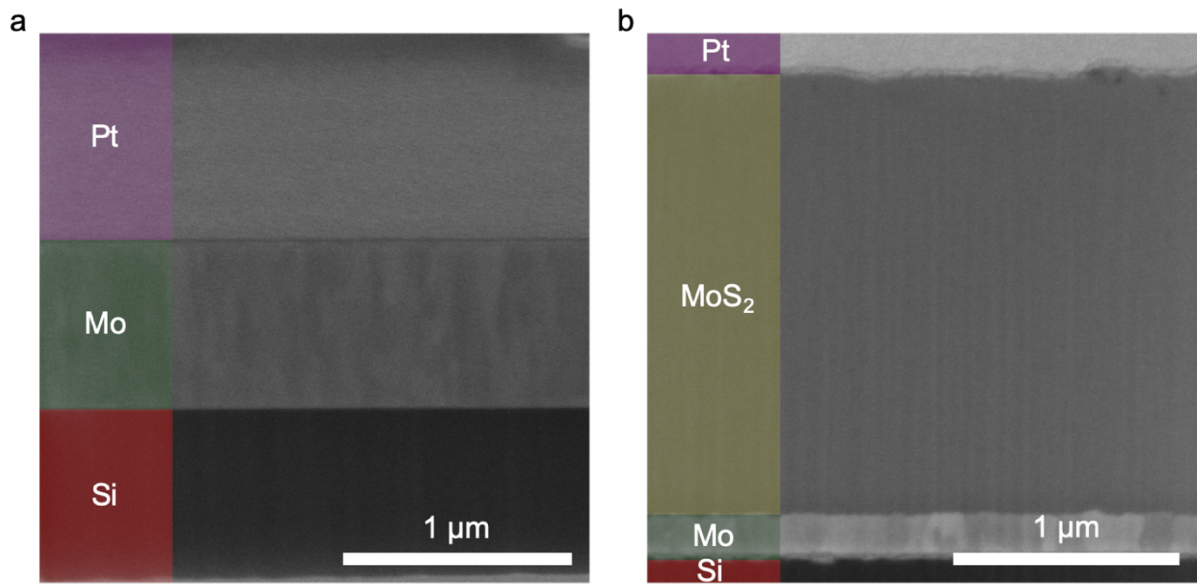

**Figure 2. (a)** and **(b)** SEM images of the FIB cross-sections corresponding to the samples shown in **Figure 1a** and **1b**. In **(a)**, the sample grown at 500 °C, we observe three distinctive regions with different contrasts, associated to the protective Pt layer (magenta), the Mo seed layer (green), and the Si substrate (red). Note that in this case the presence of $MoS_2$ is restricted to the surface of the sample. In **(b),** grown at 700 °C, the four regions are associated to the protective Pt (magenta), the $MoS_2$ layer with vertical nanosheets (yellow), the Mo seed layer (green), and the Si substrate (red).

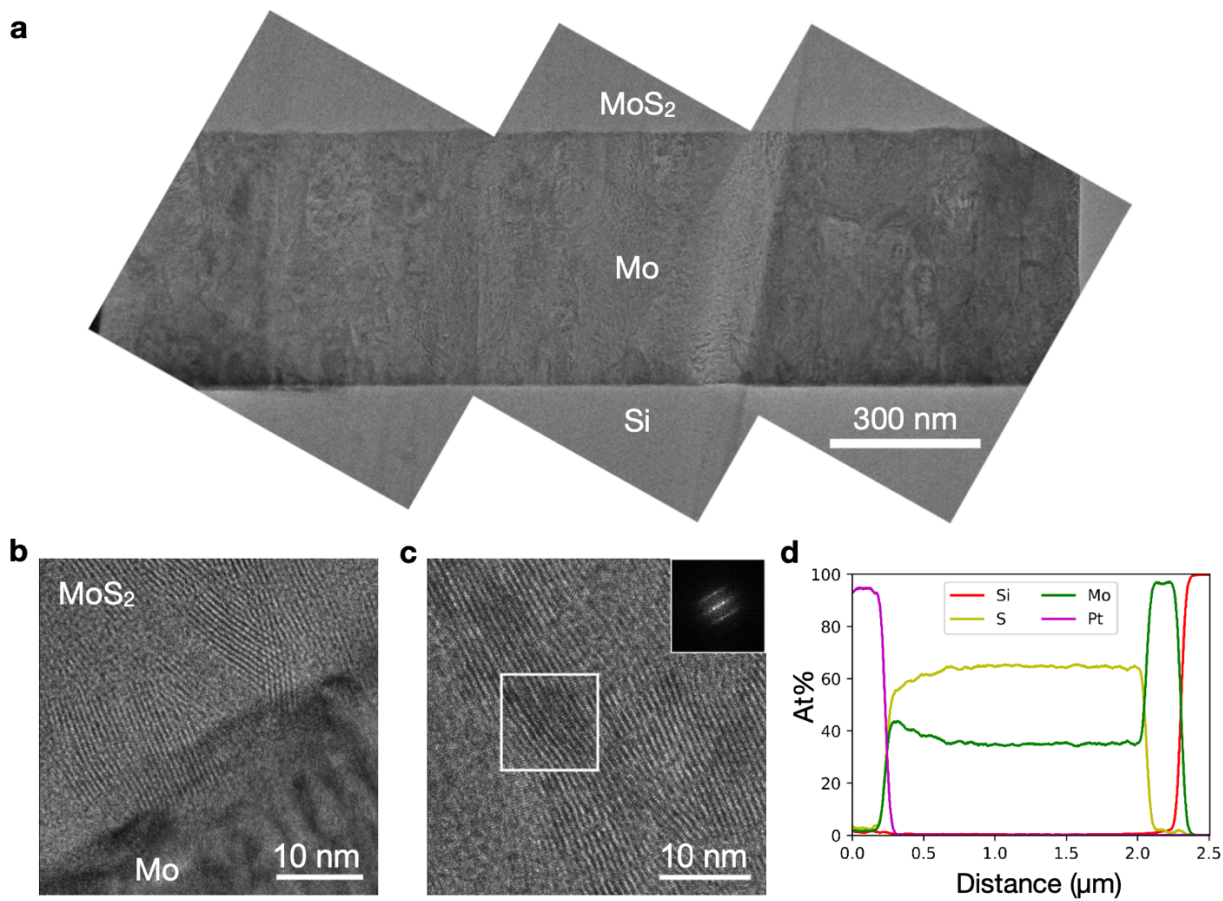

**Figure 3. (a)** Reconstructed low-magnification bright-field TEM image of a sample sulfurized at 700 °C, displaying the Mo seed layer sandwiched between v-MoS$_2$ on the top and the Si substrate on the bottom. **(b)** HRTEM image of the interface region showing how the v-MoS$_2$ nanosheets arise from the Mo-metal layer. **(c)** HRTEM image of a region containing only v-MoS$_2$ nanosheets, together with the FFT calculated in the area highlighted with a white square in the inset. **(d)** EDX line scan of a sample sulfurized for 1 hour at 700 °C indicating its elemental composition.

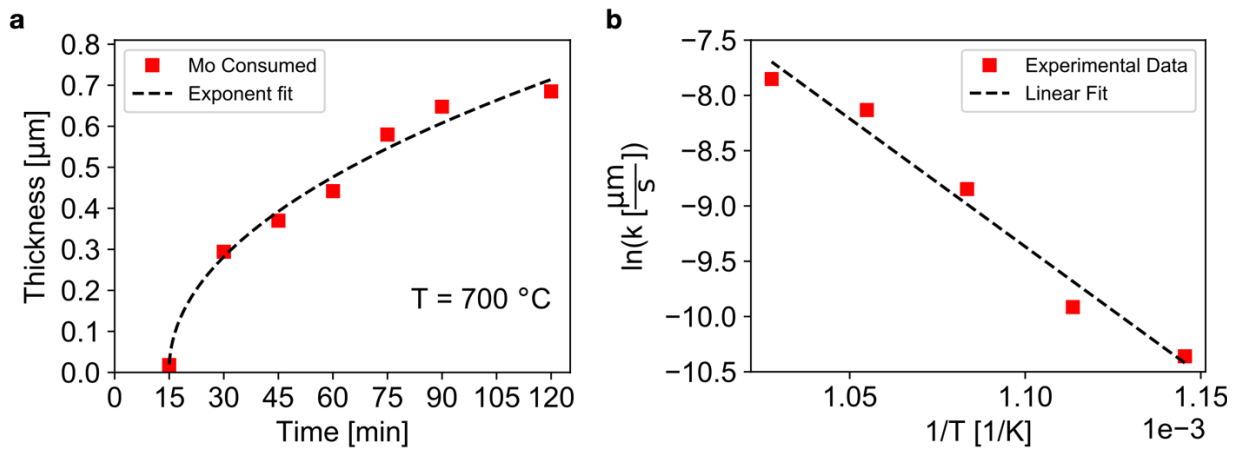

**Figure 4. (a)** The thicknesses of the consumed Mo-metal layer as a function of the reaction time for the sample grown at 700°C. The dotted line corresponds to a model fit of the form $z = K(t - t_0)^n$. The best-fit growth exponent is found to be *n*=0.48, very close to the *n*=0.5 expected for a diffusion-driven process. **(b)** An Arrhenius diagram where the logarithm of the rate constant *k* is represented as a function of the inverse of the reaction temperature. The linear fit to the data allows us to determine the activation energy required to sulfurize the Mo seed layer.

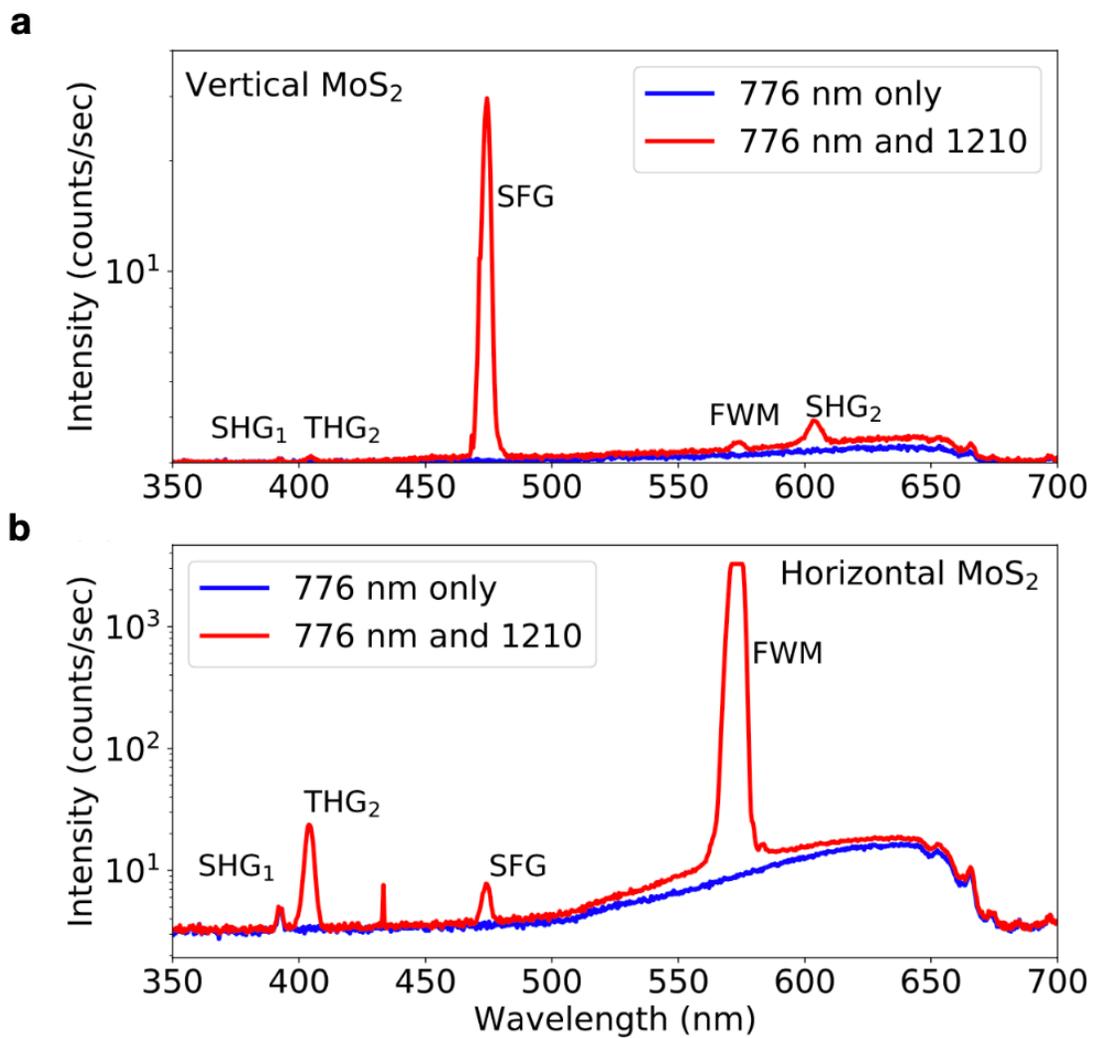

**Figure 5.** Spectra from a vertical **(a)** and horizontal **(b)** MoS$_2$ nanosheets simultaneously illuminated with ultrashort laser pulses at 776 nm and 1210 nm. The labels next to each peak indicate the corresponding nonlinear mechanisms that mediate each emission.

**ASSOCIATED CONTENT**

Additional information about the synthesis and characterization of vertically-oriented MoS$_2$ nanosheets. (PDF)

**AUTHOR INFORMATION**

**Corresponding Author**

*E-mail: s.conesaboj@tudelft.nl

**Present Addresses**

† ICTS – Centro Nacional de Microscopía Electrónica, Universidad Complutense, 28040, Spain

**ACKNOWLEDGMENT**

M.B., S.E.v.H., and S.C.-B. acknowledge financial support from ERC through the Starting Grant "TESLA" grant agreement No. 805021. M.T.R. acknowledges support from the Netherlands Organizational for Scientific Research (NWO) through the Nanofront program. The authors L.K. and J.H.-R. acknowledge funding in the framework of FP7 Ideas: European Research Council (ERC 340438-CONSTANS).

# Supporting Information

# Vertically-oriented MoS$_2$ nanosheets for nonlinear optical devices

M. Bolhuis[1], J. Hernandez-Rueda[1], S. E. van Heijst[1], M. Tinoco Rivas[1,†], L. Kuipers[1], and S. Conesa-Boj[1,*]

[1] Kavli Institute of Nanoscience, Delft University of Technology, 2628CJ Delft, The Netherlands

## A. Synthesis of vertically-oriented MoS$_2$ nanosheets.

**Figure S1** provides a schematic illustration of the growth protocol adopted for the growth of the vertically-oriented MoS$_2$ nanosheets. Before the sulfurization, a pre-treatment with Ar flow (500 sccm) during 30 min was carried out. This pre-treatment was followed by a gradually ramp up of the temperature at a rate of 10 °C /min. The temperature of zone 2, were the wafer containing the Mo seed layer was placed, was ramped up to the desired reaction temperature in the range between 500 °C and 700 °C. We note that the Ar gas flow was reduced from 500 sccm down to 150 sccm once the temperature in zone 2 reached 500 °C.

Once the temperature of zone 2 reached the target reaction temperature, we started to ramp up the temperature in zone 3, were the Sulfur powder was located, up to 220 °C at a rate of 10 °C /min. We defined the initial reaction time as the time when the temperature of the Sulfur reached 220 °C, and from this point onwards we measured the total reaction time of the Sulfur with the Mo seed layer. Once the desired reaction time was reached, the heating of both zones of the furnace was switched off at the same time. While the furnace tube cooled down to room temperature, the Ar gas flow was kept running.

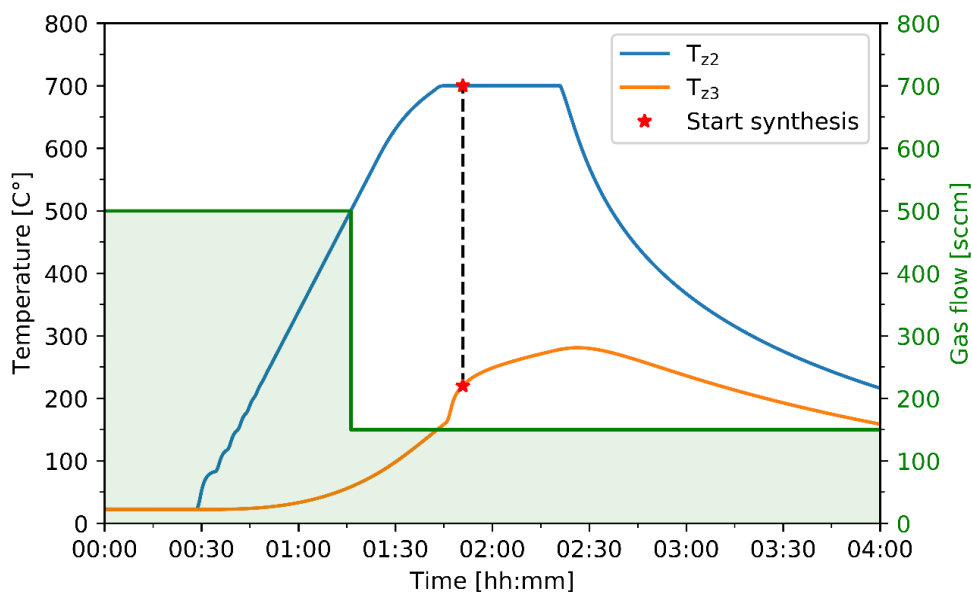

**Figure S1.** Schematic illustration of the growth protocol adopted for the growth of vertically-oriented MoS$_2$ nanosheets. We display the temperature (left *y*-axis) and the Ar gas flow (right *y*-axis) as a function of the reaction time. In this scheme, T$_{z2}$ and T$_{z3}$ indicate the temperatures of zones 2 and 3 of the furnace respectively. The stars indicate the points defined as the starting times of the reaction.

*Argon flow dependence.* – **Figure S2** displays the variation of the MoS$_2$ thickness as the Ar flow rate is increased. We found that as the Ar flow rate increased, the MoS$_2$ thickness increased as well until a maximum value of about 1 $\mu$m at an Ar flow rate of 75 sccm was reached. From this point onwards, the MoS$_2$ thickness instead decreased with further increases of the Ar flow rate. For this reason, in the design of the growth parameters, we only considered Ar flow rates above 75 sccm, for which the expected trend that higher flow rates slow the reaction and lead to thinner MoS$_2$ was verified. Specifically, for all the experiments presented in this work we fixed the Ar flow at a value of 150 sccm.[1]

It is worth mentioning that the growth of v-MoS$_2$ nanosheets via sulfurization can be strongly dependent on the Ar flow when the furnace tube is affected by flow instabilities. In **Figure S2** we also display two points (red crosses) corresponding to growths affected by such instabilities in the tube furnace. The results of these growths differed significantly from the main trend, and thus were not considered further.

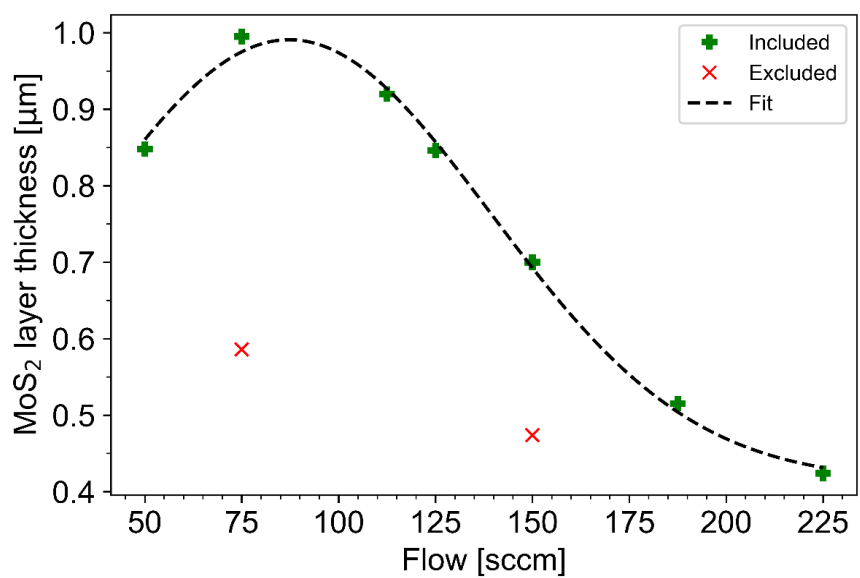

**Figure 2S.** The thickness of the grown MoS$_2$ layer as a function of the Ar gas flow. The model fit is based on the data points labelled with a green cross. We observe that for Ar flows above 75 sccm the MoS$_2$ thickness decreases monotonically as the flow is increased.

---

[1] Kong, D.; et al. Synthesis of MoS$_2$ and MoSe$_2$ films with Vertically Aligned Layers. *Nano Lett.* **2013**, 13, 1341.

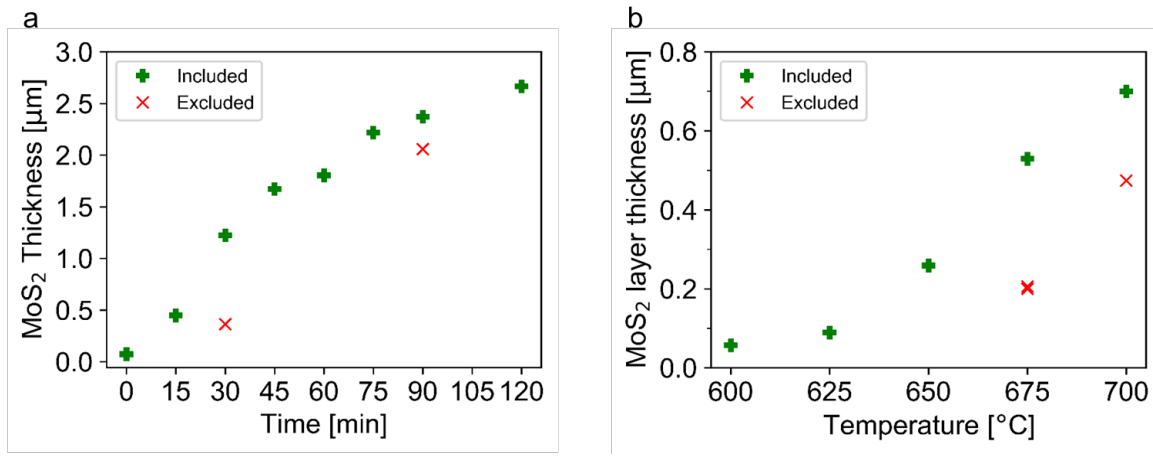

**Figure 3S.** Same as **Figure 2S** with the resulting MoS$_2$ thicknesses represented as a function of the reaction time (**3a**) and of the reaction temperature (**3b**).

In **Figures S3a** and **S3b** we display the results corresponding to the same growths reported in **Figure S2** now with the resulting MoS$_2$ thicknesses being represented as a function of the reaction time and the reaction temperature respectively. From this comparison one also observes that, while most of the data points follow the expected clear trend, the presence of instabilities in the tube furnace related to the Ar gas flow can result in significant variations in the growth rate.

**B. Structural characterization of v-MoS₂ nanosheets.**

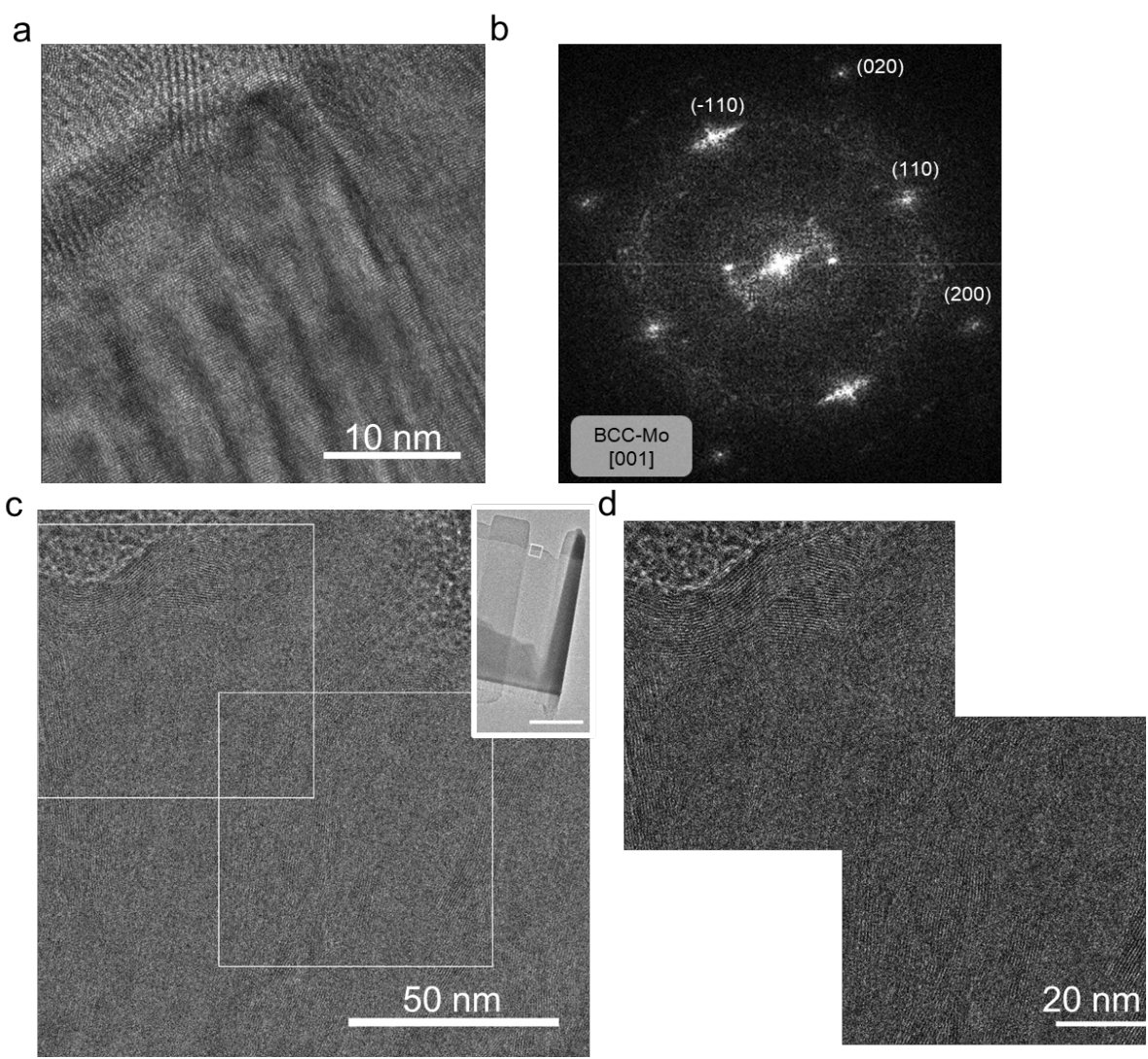

**Figure S4. (a)** and **(b)** HRTEM image taken from a representative specimen sulfurized at 700 °C displaying a region of the Mo-metal seed layer and the corresponding FFT, indicating that the Mo is characterized by the BCC crystal phase. **(c)** HRTEM image of the upper region of the grown MoS₂ (see inset) immediately below the Pt protective layer. In **(d)** we show the inverse FFT calculated from the two regions marked with white squares in **(c)**.

**Figure S4** displays a cross section HRTEM analysis of one of the representative specimens sulfurized at a temperature of 700 °C. Measurements were taken at different regions of the cross-section lamella: at the Mo seed layer (**Figure S4a**), at the interface between the Mo seed layer and the v-MoS$_2$ nanosheets (**Figure S4c**), and at the first nanometers from the surface of the sulfurized lamella (**Figure S4d**). In the first case, we also include the corresponding FFT indicating that the Mo-metal crystallizes in the BCC structure. In the latter case, the MoS$_2$ nanosheets were found to exhibit random orientations: in the first 20 nm from the surface (beneath the Pt protective layer), vertical and horizontal nanosheets appear simultaneously, as can be seen from the orientations of lattice fringes in **Figure S4d**. Afterwards, the rest of the grown MoS$_2$ is restricted to the vertical orientation.

Different scenarios have been proposed to explain the appearance of a orientation-disordered region where both horizontal and vertical MoS$_2$ are present. Our experimental findings are consistent with the theoretical modelling of the diffusion-reaction growth proposed for the synthesis of vertically-oriented MoS$_2$.[2] This model describes the v-MoS$_2$ growth from the co-existence of two different morphologies, the first an orientation-disordered region extending from the surface to the first few nanometers, and the region containing exclusively vertically-oriented MoS$_2$ nanosheets.

C. **Chemical analysis of the cross-section TEM lamellas the v-MoS$_2$ nanosheets.**

A Python analysis code was developed and used to process and interpret the EDX line spectra data. In the following we describe how the S/Mo atomic ratio was calculated at three different locations of the chemical spectra. **Figure S5a** displays the elemental atomic percentage profile taken along the cross-section lamella, from the protective layer Pt (violet line) to the Silicon wafer (red line). **Figure S5b** then shows the same raw data after applying a Savitzky-Golay filter in order to smoothen the stochastic fluctuations. **Figure S5c** represents the data restricted to the points that we defined as top and bottom of the MoS$_2$ sulfurized region. We determined the cross-point between the Pt and S (from the top) and between the S and Mo (in the bottom), and then we chose three different locations to automatically calculate the S/Mo ratio. We selected these three locations at distances of 35% (close to the surface), 50%, and 65% respectively. All the calculated S/Mo ratios were then collected and plotted in **Figure S5e.** From this figure, we can observe that for times below 15 min the sulfurization is not fully complete. From the same plot we can also observe that the S/Mo ratio remains stable along the lamella.

---

[2] Stern, C. et al. Growth Mechanism and Electronic Properties of Vertically-aligned MoS$_2$. *Scientific Reports* **2018**, 8,16480.

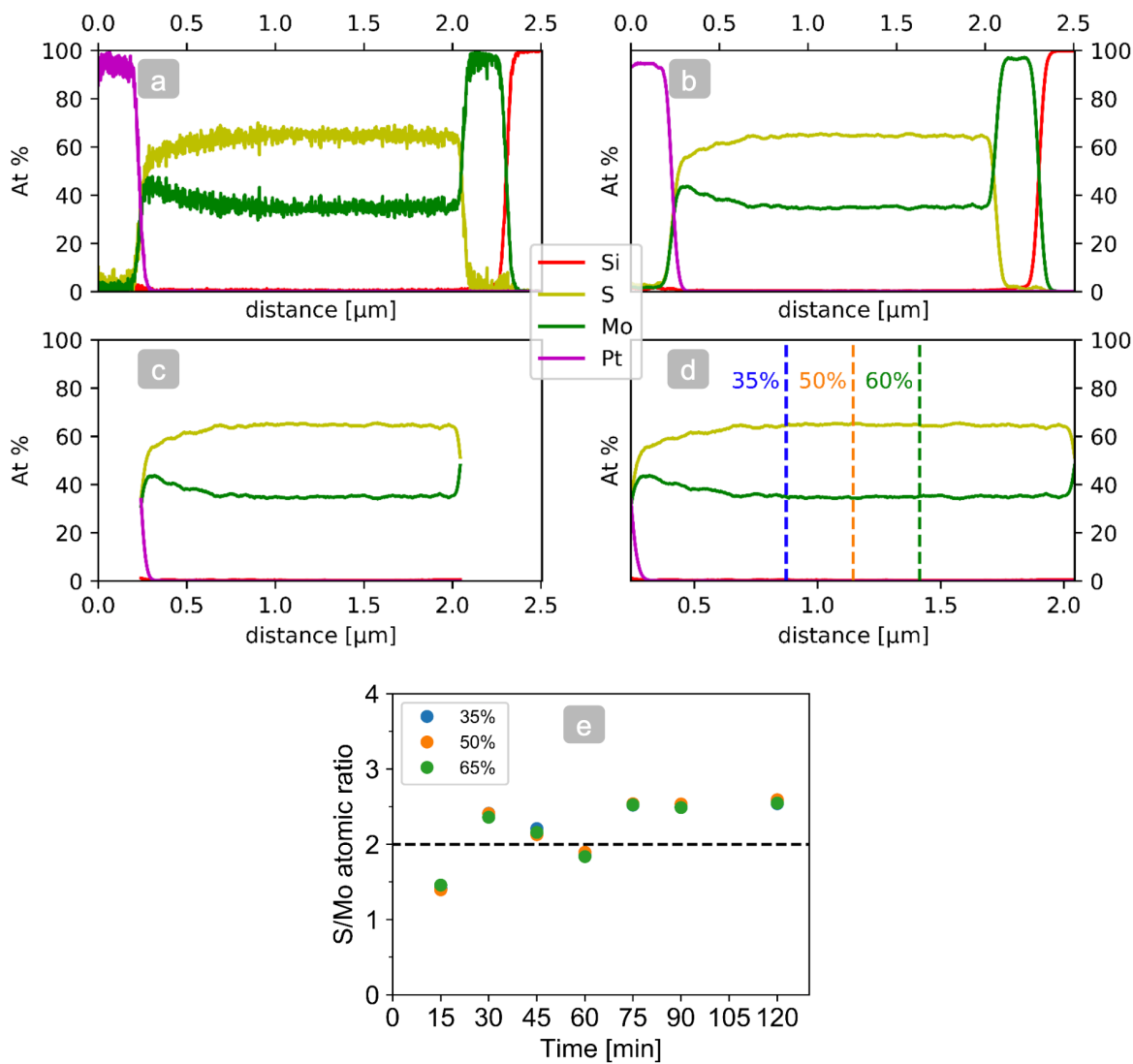

**Figure S5. (a)** The elemental atomic percentage profile taken along the cross-section lamella, from the protective layer Pt (violet line) to the Silicon wafer (red line). **(b)** Same raw data after applying a Savitzky-Golay filter in order to smoothen the stochastic fluctuations. **(c)** The same data now restricted to the points defined as the top and bottom of the $MoS_2$ sulfurized region. **(d)** The three different locations selected to calculate the S/Mo ratio, see text. **(e)** Summary of the calculated S/Mo ratios calculated as a function of the reaction time.

## D. Raman spectroscopy of exfoliated MoS$_2$ flakes.

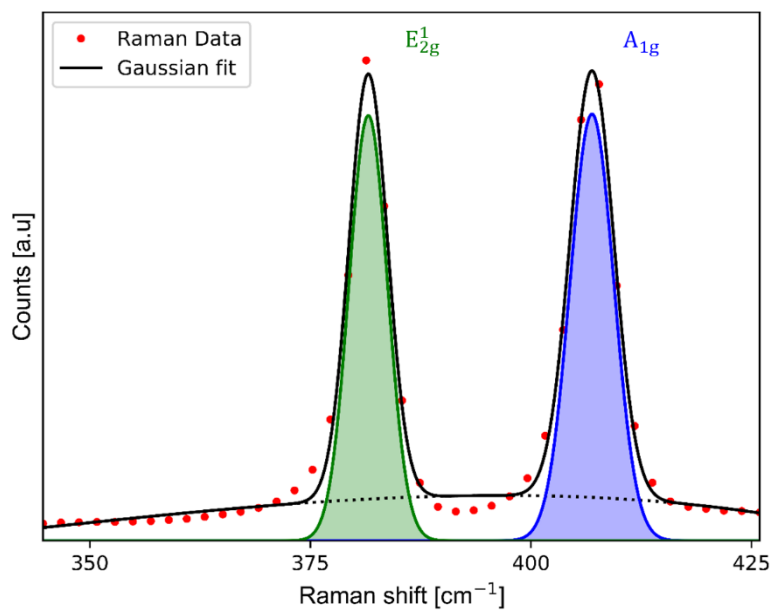

**Figure S6.** Raman emission spectra taken on a regular exfoliated MoS$_2$ flake using the same experimental set-up as in **Figures 1c** and **1d** in the main manuscript. We observe that the intensity of the two dominating Raman modes, $E^1_{2g}$ and $A_{1g}$, is very similar and therefore the ratio of intensities is close to unity. These two Raman peaks are associated to the in-plane and out-of-plane vibration modes respectively.